\documentclass{amsart}

\newtheorem{theorem}{Theorem}[section]

\newtheorem{lemma}[theorem]{Lemma}

\theoremstyle{definition}
\newtheorem{definition}[theorem]{Definition}
\newtheorem{example}{Example}
\newtheorem{remark}[theorem]{Remark}
\def\ad{\textup{ad}}
\def\Span{\textup{span}}
\def\tR{\tilde\R}
\def\gr{\textup{gr}}
\def\RX{{\R_{X}}}
\def\II{\mathcal{I}}
\def\JJ{\mathcal{J}}
\def\C{{\mathbb{C}}}
\def\Z{{\mathbb{Z}}}
\def\R{{\mathcal{R}}}

\def\D{{\mathcal{D}}}
\def\A{{\mathcal{A}}}
\def\OO{{\mathcal{O}}}
\def\RA{{\R_{\A}}}
\def\FF{{\mathcal{F}}}
\def\AA{\mathfrak{A}}
\def\VV{\mathcal{V}}
\def\V{\mathcal{V}}
\def\g{\mathfrak{g}}
\def\gOO{\g_{\OO}}
\def\gFF{\g_{\FF}}
\def\D{{\mathcal{D}}}
\def\supp{\textup{supp}\,}
\def\ord{\textup{ord}\,}
\def\Ord{\textup{Ord}\,}
\def\spec{\textup{Spec}\,}

\def\End{\textup{End}}
\def\deg{\textup{deg}\,}
\def\Deg{\textup{Deg}\,}
\def\Ann{\textup{Ann}}
\title{$\D$-modules and Darboux
Transformations}
\author{Yuri Berest}
\address{Department of Mathematics, University of California,
Berkeley}

\author{Alex Kasman}
\address{Department of Mathematics and Statistics, Concordia
University \\and\\ Centre de recherches math\'ematiques, Universit\'e de
Montr\'eal}

\begin{document}

\begin{abstract}
A method of G.~Wilson for generating commutative algebras of
ordinary differential operators is extended to higher
dimensions. Our construction, based on the theory of $\mathcal{D}$-modules,
leads to a new class of examples of \textit{commutative rings of
partial differential operators} with rational spectral 
varieties. As an application, we briefly discuss their
link to the bispectral problem and to the theory of 
lacunas.
\end{abstract}

\maketitle

\section{Introduction}

\subsection{}
The purpose of this paper is to discuss a certain algebro-geometric
construction of commutative rings of partial differential operators
with singular spectral varieties.  The principal idea of the method to
be used goes back to the work of Krichever (see, e.g., \cite{Kr} for a
review), though our effort here has been inspired by more recent
developments.  These are the study of algebraic Schr\"odinger
operators initiated by O.~Chalykh and A.~Veselov \cite{VCh1,VCh2} and
the work of G.~Wilson \cite{W1,W2} on ``bispectral'' algebras of ordinary
differential operators.

Our attempt is to merge the ideas of the above mentioned authors
within a more general setting and to explore some new instructive
examples.

\subsection{} In \cite{VCh1}--\cite{VCh2} a certain geometric
axiomatics for the Baker-Akhiezer (BA) function has been developed in
affine spaces of arbitrary dimension ($n\geq1$).  The construction
involves a finite set $\A$ of linear (homogeneous) hyperplanes in
$\C^n$ with prescribed (integer) multiplicities.  These data are
postulated to define a ``pole divisor'' of an associated BA function.
When it exists, such a function is necessarily unique and solves a
(formal) spectral problem for a certain \textit{commutative} ring
$\RA$ of partial differential operators containing a second order
(Schr\"odinger) operator.  Algebraically, the ring $\RA$ turns out to
be \textit{supercomplete} in the sense that the minimal number of generators
of $\RA$ (as a $\C$-algebra) exceeds its Krull dimension, i.e. the
dimension of the spectrum $\spec\RA$, while geometrically the latter
proves to be a singular variety.  Within the ``rational'' version of this
axiomatics, the differential operators in $\RA$ have \textit{constant}
principal symbols and rational lower order coefficients with
singularities located on the hyperplanes of $\A$.  These operators
possess a number of interesting analytic properties and applications
among which one should mention the link to Huygens' principle and the
theory of lacunas (see
\cite{BV1,BV2,Ber3,Ber4}).

The principal problem within the Veselov-Chalykh approach is to
determine all possible hyperplane arrangements for which a
nontrivial BA function does actually exist.  Up to now, this problem
has been completely settled in dimension $n=2$ (see \cite{Ber1,Ber2}),
while in higher dimensions ($n\geq3$) only partial results and examples
are available (see \cite{VChS,VChF1,VChF2}).

\subsection{} One may pose a more general question (cf. \cite{Kr},
Section 4): {What algebraic (or analytic) sets in $\C^{n}$ are
admissible as a singularity locus for the BA function associated to a
(non-trivial) supercomplete commutative ring of partial differential
operators?}

In this paper we  give a partial answer to this question in the
``rational'' (algebraic) situation.  Let $X$ be any open quasi-affine
algebraic variety in $\C^n$, i.e. $X=\C^n\backslash \tau^{-1}(0)$ with
some $\tau\in \C[x_1,\ldots,x_n]$, and let $\OO(X)$ and $\D(X)$
stand for the rings of regular functions and differential
operators on $X$ respectively.
It is a consequence of Theorem~\ref{thm:multfree} below that
there exists a nontrivial supercomplete
commutative ring $\RX\subset \D(X)$ of differential operators on $X$
whose common eigenfunction $\psi=\psi(x,\xi)$ is unique (up to a
normalization) and quasi-regular, i.e. has the form
$\psi(x,\xi):=P(x,\xi)e^{(x,\xi)}$ with $P\in
\OO(T^*X)\cong \OO(X)\otimes\C[\xi]$.

This result implies that the BA function (regarded as a function of
$x$ only) can admit singularities along \textit{arbitrary} algebraic
hypersurfaces in $\C^n$.  In Section~\ref{sec:highdim} we will give a
simple and constructive proof of this observation.  A number of
explicit examples will be treated in Section~\ref{sec:examples}.

It is worth mentioning that the ring $\RX$ does not necessarily contain
a second order operator and, hence, our examples go beyond the framework
of the Veselov-Chalykh  axiomatics; however, as we will see in
Section~\ref{sec:examples}, they have equally interesting analytic
applications.

Unlike the BA functions that arise from the Veselov-Chalykh
construction, a generic $\psi$-function associated with a ring $\RX$
is \textit{not self-dual}.  Nevertheless, we demonstrate the existence
of a complete ring of partial differential operators in the
\textit{spectral parameter} having $\psi$ as a common eigenfunction.
In this way, we get an example of a nontrivial bispectral involution
in the case of several variables.

\subsection{} For ordinary differential operators, 
a \textit{complete} answer to the question stated above is a simple
consequence of a method of G.~Wilson \cite{W1}.  His technique was
reformulated later in terms of Darboux transformations in
\cite{BHY1,BHY2} and \cite{K1,KR}.  The method used in the present
work can be regarded as a further generalization of this approach with
a view to address the problem for
\textit{partial} differential operators ($n>1$).  A natural framework
for such a generalization is provided by \textit{the theory of
$\D$-modules}.  Our intention is to demonstrate that the language of
$\D$-modules is quite relevant and illuminating in this context.  It
brings some new insight even in the case $n=1$ (see
Section~\ref{sec:n=1}).  In this paper we focus mostly on the
construction of families of commuting differential operators rather
than on the study of their properties.  In particular, we only briefly
discuss the bispectral problem which was one of our original
motivations.  We believe, however, that the understanding of this
phenomenon within the theory of $\D$-modules is a promising problem
which deserves a separate and more detailed investigation.

\section{The theory of $\D$-modules: basic concepts}
We start with a brief summary of some results and definitions from the
theory of $\D$-modules. Our consideration will be restricted to the
purely global (algebraic) case.  For a more general and
comprehensive treatment of the subject, the reader is referred to the
literature \cite{Bjo,Bern1,Meb,Mal1} (see also \cite{Mal2,Mais}).

\subsection{Regular Differential Operators}\label{sec:regdifops}
Let $X$ be a non-singular irreducible affine algebraic variety over
$\C$.  We may (and will) regard $X$ as being imbedded in the standard
$n$-dimensional affine space $\C^n$, i.e. we identify $X$ with a set
of common zeros of a (prime) ideal $\II(X)$ in the polynomial ring
$\C[x]$.  Let $\OO(X)$ be an \textit{affine algebra} associated with
$X$ (the coordinate ring of \textit{regular} functions on $X$).  By
our identification, we have $\OO(X)\cong\C[x]/\II(X)$. Then, $\OO(X)$
is an integral domain.  We write $\FF(X)$ for the quotient field of
$\OO(X)$.

Let $\gFF(X)$ be the set of all $\C$-linear derivations on
$\FF(X)$ and let $\gOO(X)\subset\gFF(X)$ be its proper subset that
preserves the subring $\OO(X)\subset\FF(X)$, i.e.
$\gOO(X):=\{\partial \in \gFF(X)\ |\ \partial\in\End_{\C}(\OO(X))\}$.

\begin{definition}
The \textit{ring of regular (algebraic) differential operators} on $X$
is the ring of $\C$-linear endomorphisms of $\OO(X)$ generated by the
derivation operators from $\gOO(X)$ and the multiplication operators
from $\OO(X)$ itself.  This ring is denoted by $\D(X)$.
\end{definition}

The following examples will be of basic concern in the present work.
\begin{example}
If $X=\C^n$ then $\D(X)$ is the \textit{Weyl algebra}
$A_n(\C)\cong\C\langle x,\frac{\partial}{\partial x}\rangle$, i.e. the algebra of
differential operators with polynomial coefficients.
\end{example}
\begin{example}\label{examp:2}
Let $X$ be any smooth algebraic variety.  For $\tau\in\OO(X)$,
${1}/{\tau}$ is defined as a regular function on the complement
$X_{\tau}:=X\backslash\tau^{-1}(0)$ of the zero set of $\tau$ in $X$.  We
write $\OO(X)_{\tau}$ for the algebra generated by $\OO(X)$ and
${1}/{\tau}$ and call it the \textit{localization} of the ring
$\OO(X)$ at $\tau$.  Similarly, the ring $\D(X_{\tau})$ of regular
differential operators on $X_{\tau}$ can be identified with a
\textit{localization} of $\D(X)$ at $\tau$, i.e.
\begin{equation}
\D(X_{\tau})\cong \D(X)_{\tau}:=\OO(X)_{\tau}\otimes_{\OO(X)}\D(X).
\end{equation}
In particular, if $X=\C^n$ then $X_{\tau}=\C^n\backslash
\tau^{-1}(0)$, where $\tau\in\C[x]$, is an open \hbox{(quasi-)affine}
variety of dimension $n$ and the ring $\D(X_{\tau})$ is identified
with the \textit{localized Weyl algebra}
$A_n[\tau^{-1}]:=\C[x,\tau^{-1}]\otimes_{\C[x]}A_n(\C)$.
\end{example}

\subsection{$\D$-modules}
In general, the ring $\D(X)$ carries a natural filtration
$\OO(X)\equiv \D_0(X)\subset\D_1(X)\subset\D_2(X)\subset\ldots$ provided by
orders of differential operators (so that $\D_k$ is the subspace of operators
of order $\leq k$) with $[\D_k,\D_l]\subset \D_{k+l-1}$ and
$\D(X)=\cup_{k\geq0} \D_k(X)$.  The \textit{associated graded ring}
$\gr(\D(X)):=\oplus_{k\geq0} \D_k/\D_{k-1}$ is a commutative noetherian
ring isomorphic to the affine algebra of the cotangent bundle of $X$,
i.e. $\gr(\D(X))\cong\OO(T^*X)$.  There is a natural mapping
$\sigma:\D(X)\to \gr(\D(X))$ which associates to a differential
operator $Q\in\D(X)$ its \textit{principal symbol} $\sigma(Q)$, a
polynomial function on $T^*X$ homogeneous along the
fibers of $T^*X$.  The following properties are immediate:
(a) $\sigma(Q_1Q_2)=\sigma(Q_1)\sigma(Q_2)$;
(b) if ${\mathfrak{A}}\subset \D(X)$ is a left ideal in $\D(X)$
then $\sigma({\mathfrak{A}})$ is an ideal in $\gr(\D(X))$; moreover, 
(c) if $Q_1,\ldots,Q_l\in{\mathfrak{A}}$ can be chosen in such a way that
$\sigma(Q_1),\ldots,\sigma(Q_l)$ generate $\sigma({\mathfrak{A}})$ in
$\gr(\D(X))$ then $Q_1,\ldots,Q_l$ generate ${\mathfrak{A}}$ in $\D(X)$.
As a consequence, we note that $\D(x)$ is a left (right) noetherian ring.

Let $M$ be a left unitary $\D(X)$-module.  A $\D$-\textit{filtration}
on $M$ is a sequence $M_0\subseteq M_1\subseteq \ldots$ of
$\OO(X)$-submodules such that (i) $\cup_{k\geq 0} M_k=M$ and (ii)
$\D_k M_l\subseteq M_{k+l}$ for all $k,l\in\Z_+$.  If the last inclusion
(ii) holds with an equality for all sufficiently large $l\in \Z_+$
then the $\D$-filtration is said to be \textit{good}.  In fact, $M$
can be equipped with a good $\D$-filtration if and only if it is
finite over $\D(X)$.  Given a system of generators
$\{e_1,\ldots,e_s\}$ of a finite $\D$-module, one can define a good
$\D$-filtration by setting $M_k:=\D_k{\langle}e_1,\ldots,e_s\rangle$
for every $k\in\Z_+$.  Such $\D$-filtrations are called \textit{standard},
any two of them, say $\{M_k\}$ and $\{M'_k\}$, being equivalent in the
sense that $M_k\subseteq M'_{k+k_0}$ and $M'_k\subseteq M_{k+k_0}$ for
some $k_0\in\Z_+$ and for all $k\in \Z_+$.

The important geometric invariant of a $\D$-module $M$ is its
\textit{characteristic variety} $V_M$.  Let, for simplicity,
$X=\C^n$, and let $M$ be a finite module on $X$ over the Weyl algebra
$\D=A_n(\C)$.  Consider the graded $\gr(\D)$-module
$\gr_{\D}(M):=\oplus_{k\geq 0}M_k/M_{k-1}$ associated with some good
$\D$-filtration on $M$ and take its left annihilating ideal in
$\gr(\D)$:
\begin{equation}
I_{\D}(M):=\{\sigma\in \gr(\D):\sigma\cdot \gr_{\D}(M)=0\}.
\end{equation}
The variety $V_M$ is then defined as the zero set of $I_{\D}(M)$ in
$T^*X\cong \C^{2n}$, i.e.
\begin{equation}
V_M:=\{(x,\xi)\in \C^{2n}:\sigma(x,\xi)=0\ \textup{for\ all}\
\sigma\in I_{\D}(M)\}.
\end{equation}
Clearly, $V_M$ is a $\xi$-conic algebraic variety in $\C^{2n}$ whose
``vanishing'' ideal $\JJ_{M}$ is equal to the radical of $I_{\D}(M)$
(Hilbert's Nullstellensatz) and called the \textit{characteristic
ideal} of $M$.  It turns out that both $V_M$ and $\JJ_M$ depend only
on $M$ and not on the filtration chosen.

If $\pi:T^*X\to X$ is the canonical projection, one defines the
\textit{support} of a $\D$-module $M$ as a $\pi$-image of its
characteristic variety
\begin{equation}
\supp(M):=\pi(V_M)=\{x\in\C^n:\exists (x,\xi)\in V_M\}.
\end{equation}
The dimension of $V_M$ is called the \textit{Bernstein dimension}
$d(M)$ of $M$.  It is a fundamental fact in the theory
(cf. \cite{Bern1,Bern2}) that $d(M)\geq n$ for every nonzero $M$.
This suggests that one should single out the class of so-called
\textit{holonomic} $\D$-modules ${\mathcal{B}}_n$ (the \textit{Bernstein class})
for which $d(M)=n$.  Holonomic $\D$-modules possess a number of
finiteness properties that make them of particular interest from the
analytic point of view (see, e.g., \cite{Bjo}).  For example, every
$M\in {\mathcal{B}}_n$ has  finite length (i.e., any ascending chain of
$\D$-submodules in $M$ becomes stationary).  In view of Stafford's
theorem, this implies the cyclicity of $M$ since $\D$ is a simple ring
of infinite length.  As a whole, the Bernstein class ${\mathcal{B}}_n$ is stable
under certain \textit{basic operations} on $\D$-modules such as
taking direct-inverse images, localization and the Fourier
transform.

In general, the precise determination of a characteristic variety is a
difficult problem, even in the case of holonomic $\D$-modules.  On the
other hand, if a $\D$-module $M$ has a characteristic variety $V_M$ of
a particularly simple structure one can expect to describe it
explicitly.  In Section~\ref{sec:n=1}, we illustrate this
remark in the \textit{simplest} case: for holonomic $\D$-modules $M\in{\mathcal{B}}_1$ supported at a finite set of points in the complex plane,
$\supp(M)=\{\lambda_1,\ldots,\lambda_s\}$.  Using the basic operations
on $\D$-modules we then recover the Wilson construction of rank one
commutative algebras of ordinary differential operators.  The very
existence of such algebras reflects a certain symmetry (or, to be more
precise, ``simplicity'') of the
underlying $\D$-module and its characteristic variety.

\section{Holonomic $\D$-modules and Commutative Algebras of Ordinary
Differential Operators}\label{sec:n=1}

In this section we review a construction of rank one commutative rings of
differential operators in dimension $n=1$ due to G.~Wilson \cite{W1}.
Our purpose is to analyze Wilson's scheme in terms of holonomic
$\D$-modules over the Weyl algebra and its
localizations. Reinterpreting Wilson's construction in these terms
motivates the higher dimensional generalization to be presented in
Section~\ref{sec:highdim}.

\subsection{$\D$-modules with point support} Let $X=\C^n$ and let $M$ be a
finite left $\D$-module on $X$ with $\D\cong A_n(\C)$.  Suppose that
$\supp(M)=\{\lambda\}$, $\lambda\in\C^n$.  In this case one can
recover the structure of $M$ explicitly.  Indeed, it is clear that
$\dim_{\C} V_M\leq n$, where $V_M$ is the characteristic variety of
$M$.  In view of Bernstein's inequality this implies $d(M)=n$ and
$V_M=T^*_{\lambda} \C^n$.
Hence, $M$ is a holonomic $\D$-module, $M\in{\mathcal{B}}_n$.  If
$M_0\subset M_1\subset M_2\subset\ldots$ is a good filtration on $M$,
the associated graded module $\gr_{\D}M$ is supported at $\lambda$.
By the Nullstellensatz one has 
\begin{equation}
{\mathfrak{m}}_{\lambda}^N\sigma(a)=0\label{eqn:one}
\end{equation}
with some $N\in\Z_+$ for any
$\sigma(a)\in\gr_{\D}(M)$, where ${\mathfrak{m}}_{\lambda}\equiv \C[x]\langle x-\lambda\rangle$ is the
(maximal) ideal in $\C[x]$ corresponding to the point $\lambda$.
But then, \eqref{eqn:one} is also true for any element in $M$:
${\mathfrak{m}}_{\lambda}^Na=0$.
To clarify the structure of $M$ we start with the simplest example:
$\Delta(\lambda):=\D/\D{\mathfrak{m}}_{\lambda}$.  Clearly,
$\supp\Delta(\lambda)=\{\lambda\}$ since ${\mathfrak{m}}_{\lambda}^{N+1}
Q\in \Delta(\lambda)$ for every $Q\in\D$ of order $\ord Q\leq N$.
Denote by $\delta_{\lambda}(x)$ the canonical generator of
$\Delta(\lambda)$: $\delta_{\lambda}(x):=1+\D{\mathfrak{m}}_{\lambda}$ (so
$\delta_{\lambda}(x)$ is the image of $1\in\D$ under the natural
surjection $\D\to\Delta(\lambda)$).  It is easy to verify that 
$$
\Delta(\lambda)\cong
\D\langle\delta_{\lambda}(x)\rangle\cong\C[\partial]\langle\delta_{\lambda}(x)\rangle,
$$
where $\C[\partial]$ is the ring of differential operators with
constant coefficients in $\C^n$.

The following lemma determines the algebraic structure of a finite
$\D$-module with point support.

\begin{lemma}\label{lem:simple} If $M$ is a finite-type left $\D$-module on $X=\C^n$
with $\supp M=\{\lambda\}$, $\lambda\in\C^n$, then $M$ is isomorphic
to a finite direct sum of copies of $\Delta(\lambda)=\D/\D
{\mathfrak{m}}_{\lambda}$.  In particular, $\Delta(\lambda)$ is a simple
$\D$-module (i.e. has no nontrivial submodules over $\D$).
\end{lemma}

For a proof of this result the reader is referred to \cite{Bjo},
Ch. 5 (see also \cite{Mal2,Mais}).

\subsection{Left Ideals of the Weyl Algebra}

Now let $n=1$ and let $M$ be any finite left $\D$-module
over the Weyl algebra $\D=A_1(\C)$.  We choose a system of generators
$\{e_1,\ldots,e_s\}$, $M=\D\langle e_1,\ldots,e_s\rangle$, and fix the
standard filtration $M_1\subset M_1\subset M_2\subset\ldots$ on $M$
associated with this system $M_k:=\D_k\langle e_1,\ldots,e_s\rangle$.
Then, we denote by ${\AA}={\AA}(e_1,\ldots,e_s)\subset\D$
the left ideal in $\D$ that annihilates the elements
$\{e_1,\ldots,e_s\}$
\begin{equation}
{\AA}:=\Ann_{\D}(e_1,\ldots,e_s)=\{ Q(x,\partial)\in\D\ :\
Q(x,\partial) e_j=0,\ j=1,\ldots,s\}.\label{eqn:two}
\end{equation}
Put ${\AA}_k:=\D_k\cap {\AA}$ and consider the associated
graded ideal in $\gr(\D)$:
$$\gr_{\D}({\AA})=\bigoplus_{k\geq 0}
\AA_k/\AA_{k-1}$$  containing the principal
symbols of the differential operators in ${\AA}$.
Clearly, $\gr_{\D}({\AA})$ is a homogeneous ideal in the
polynomial ring $\C[x,\xi]\cong\gr(\D)$, and therefore it can be
presented in the form
\begin{equation}
\gr_{\D}({\AA})=\bigoplus_{k\geq0} I_k({\AA})\xi^k.
\end{equation}
Here, $I_0({\AA})\subset I_1({\AA})\subset
I_2({\AA})\ldots$ is an \textit{increasing} sequence of ideals in
$\C[x]$ that corresponds to leading terms of the differential
operators in ${\AA}$.  Since $\C[x]$ is a principal ideal domain
(recall, $n=1$), each $I_k({\AA})$ is cyclic while
the sequence $\{I_k\}$ stabilizes: $I_q=I_{q+1}=I_{q+2}=\ldots$
starting from some $q\geq 0$.  We write $\tau_k=\tau_k(x)$ for the
generators of $I_k({\AA})$ in $\C[x]$, and notice that each
$\tau_k$ is determined uniquely (up to a constant factor).  Let
$$p:=\textup{min}\,\{k\in\Z_+\ :\ I_k({\AA})\not=0\}$$
\textup{and}
$$
  q:=\textup{min}\,\{k\in\Z_+\ :\ I_k=I_{l}\ \forall l\geq k\}
$$
so that $q\geq p$.  Then, the polynomials
$\{\tau_p(x)\xi^p,\ldots,\tau_q(x)\xi^q\}$ generate the graded ideal
$\gr_{\D}({\AA})$ in $\C[x,\xi]$.  We refer to the integers $p$ and
$\alpha_q:=\deg \tau_q(x)$ as the \textit{order} and \textit{degree}
of the ideal ${\AA}$ respectively, and write
$\Ord{\AA}:=p$, $\Deg{\AA}:=\alpha_q$.  A set of
differential operators
$\{L_p(x,\partial),\ldots,L_q(x,\partial)\}\subset{\AA}$ such
that $\sigma(L_r)=\tau_r(x)\xi^r$, $r=p,p+1,\ldots,q$, is called a
\textit{standard basis} of the ideal ${\AA}$ (cf. \cite{BrM}).
The differential operator $L_p$ of the minimal order in a standard
basis is determined uniquely, once a normalization of its leading
coefficient is fixed.  If follows from our previous discussion that a
standard basis is indeed a system of generators of the ideal
${\AA}$. In fact, it is not difficult to prove that
${\AA}=\D\langle L_p\rangle+\D\langle L_q\rangle$ in agreement
with a well-known result of Stafford that every ideal in the Weyl
algebra $A_n(\C)$ can be generated by two elements (see \cite{Bjo},
Ch.~I).

\subsection{Wilson's Construction}
Let $M=\oplus_{j=1}^N \Delta(\lambda_j)$ be a direct sum of finite
left $\D$-modules with point support.  It is easy to see that
$V_M=\cup T^*_{\lambda_j}\C$ and, hence, that $\supp
M=\{\lambda_1,\ldots,\lambda_N\}$.  We fix a $\C$-linear
finite-dimensional vector space $\VV\subset M$, \textit{homogeneous}
with respect to the direct decomposition above: $\VV=\oplus_j \Delta(\lambda_j)\cap
\VV$, and such that $M=\D\langle \VV\rangle$.  Then, according to
Lemma~\ref{lem:simple}, $\VV$ has a finite basis of the form $e_j\cong
p_j(\partial)+\D\mathfrak{m}_{\lambda_j}$.  Following Wilson
\cite{W1}, we also consider the subring $\R_{\VV}\subset \C[x]$ of
polynomials $P\in\C[x]$ with the property $P.\VV\subset \VV$, i.e.
$P.\in\End_{\C}(\VV)$.  Since $\supp M$ is finite, $\R_{\VV}$ is not
empty.  In particular, it contains all $P\in\cap_j
{\mathfrak{m}}_{\lambda_j}^k$ for a sufficiently large $k\in\Z_+$.
The left ideal ${\AA}:=\Ann_{\D}(\VV)$ annihilating $\VV$ has the
structure of a \textit{right} module over $\R_{\VV}$.  Indeed,
$({\AA}. P)\VV=\AA(P.\VV)\subset {\AA} \VV=0$ so that ${\AA}\supset
{\AA}.P$ by definition of ${\AA}$.  Note also that $\Ord {\AA}=0$.

Now we apply the operation of a \textit{Fourier transform} $M\to
M^{\vee}$ to the $\D$-module $M$.  By definition, this is a (unique)
bijection induced by the formal involution on the Weyl algebra
$\D=A_1(\C)$: $x\mapsto-\partial_x$, $\partial_x\mapsto x$.  One can
prove \cite{Bern1} that $M^{\vee}\in{\mathcal{B}}_1$ provided
$M\in{\mathcal{B}}_1$.  If $\VV^{\vee}\subset M^{\vee}$ is the Fourier
image of $\VV\subset M$, then $\VV^{\vee}$ is annihilated by
${\AA}^{\vee}$, the image of ${\AA}$ under the Fourier
involution in $\D$.  The fact that $\Ord{\AA}=0$ implies that
${\AA}^{\vee}\cap\C[\partial]$ is not empty.  In particular,
$\Deg{\AA}^{\vee}=0$.  

Let $\{G_p(x,\partial),\ldots,G_q(x,\partial)\}$ be a standard basis
of the ideal ${\AA}^{\vee}$ so that
${\AA}^{\vee}=\D\langle G_p\rangle + \D\langle G_q\rangle$.
Clearly, we may choose $G_q\in\C[\partial]$, while
$G_p(x,\partial)\not\in\C[\partial]$ in view of the
\textit{homogeneity} of the space $\VV$.  Recall  that the operator
$G_p$ is determined \textit{uniquely} by the ideal ${\AA}^{\vee}$ (up
to an inessential constant factor).  We call the leading term of this
operator $\tau=\tau_p(x)$ the \textit{$\tau$-function} associated with
the ideal ${\AA}^{\vee}$.  The idea now is to localize the Weyl
algebra $A_1(\C)$ at $\tau=\tau_p(x)$ and to consider the
\textit{extension} of the ideal ${\AA}^{\vee}$ in this localization.

Thus, we set
$$\D_{\tau}:=\D[\tau^{-1}]\cong\C[x,\tau^{-1}]\otimes_{\C[x]} A_1(\C)$$
and denote by ${\AA}_{\tau}^{\vee}$ the left ideal in $\D_{\tau}$
generated by ${\AA}^{\vee}$.  It follows that
${\AA}_{\tau}^{\vee}$ is a (principal) ideal with a
unique generator $D=D(x,\partial):=\tau(x)^{-1}\circ G_p(x,\partial)\in
\D_{\tau}$, since $\ker G_p(x,\partial)\subset \ker G_q(\partial)$ by
construction.  On the other hand, ${\AA}_{\tau}^{\vee}$ has the
structure of a \textit{right} module over the (commutative) ring
$\R_{\VV}^{\vee}\subset \C[\partial]$, the Fourier transform of
$\R_{\VV}\subset\C[x]$.  Hence, we conclude that for every $L_0\in
\R_{\VV}^{\vee}$ there exists an operator $L\in\D_{\tau}$ such that
$D\circ L_0=L\circ D$ in ${\AA}_{\tau}^{\vee}$.  The family of
all differential operators $L\in\D_{\tau}$ obtained in this way forms a
\textit{commutative} ring in the algebra $\D(X)$ of regular
differential operators on $X=\C^1\backslash \tau^{-1}(0)$, isomorphic
to $\R_{\VV}$.  It is not difficult to show (cf., \cite{W1}) that $\spec
\R_{\VV}$ is a rational curve with cuspidal singularities located
at the points of $\supp M$.  In this way, we recover Wilson's scheme of
rational Darboux transformations in dimension $n=1$ (cf.
\cite{BHY1,KR}).

\section{Algebraic Darboux Transformations for Partial Differential
Operators}\label{sec:highdim}

The purpose of this section is to generalize Wilson's construction to
higher dimensions.  We propose a procedure for generating
(super)complete commutative rings of partial differential operators in
arbitrary dimension whose common (formal) eigenfunction
$\psi=\psi(x,\xi)$ (being necessarily unique) has a particularly
simple structure.  By analogy with the one-dimensional case, we call
this procedure the \textit{algebraic Darboux transformation}.  The
fact that, in general, $\psi(x,\xi)$ has no trivial symmetry between
spatial and spectral variables leads to a new example of the
\textit{bispectral duality} (in the sense of \cite{DGr,Gr1,W1}) in the case
of several variables.  More precisely, we extend the algebraic Darboux
transformation to its ``bispectral'' counterpart and construct a
commutative ring of (partial) differential operators in the spectral
parameter sharing the same eigenfunction $\psi=\psi(x,\xi)$.  However,
unlike the one dimensional case, we are far from claiming that our
procedure completely settles the bispectral problem in many
dimensions.  Our intention is to explore some first instructive
examples in this direction.  (For some other examples, see \cite{BerG}.)

\subsection{Commuting differential operators associated
with $\D$-modules supported by an algebraic hypersurface}

\subsubsection{The multiplicity free case}\label{sec:multfree}  Let $q(x)\in\C[x]$
 be an irreducible polynomial in $n\geq1$ variables, $x:=(x_1,\ldots,x_n)$,
and let $\mathfrak{m}_q$ stand for the corresponding prime ideal in
the polynomial ring.  Extending the procedure developed in
Section~\ref{sec:n=1}, we start with the $\D$-module
$\Delta(q):=\D/\D
\mathfrak{m}_q$ on $\C^n$ supported at the zero set of $q$.  We single
out a cyclic
submodule $M:=\D\langle e \rangle$ of $\Delta(q)$ generated by the element
$e:=\tau(-\partial)+\D \mathfrak{m}_q$ where
$\tau=\tau(x)$ is an arbitrary fixed polynomial in $\C[x]$, and write
$\AA:=\Ann_{\D}(e)$ for the left annihilator of $e$ in $\D$.  Clearly,
$\AA$ is a left ideal in the Weyl algebra of zero order,
$\Ord\AA=0$, since $\mathfrak{m}_q^N e=0$ in $\Delta(q)$ for all
sufficiently large $N\in\Z_+$ ($N\geq \deg\tau+1$).

Let
$\tR_e$ be the subring of differential operators in $\D$ such that
$$\tR_e:=\{Q\in\D\ :\ e\in\ker_M(Q-\lambda)\ \textup{for some}\
\lambda\in\C\}.$$
Put $\R_e:=\tR_e\cap\C[x]$.  Then, $\R_e$ is \textit{not empty} since
$\AA\subset \tR_e$ while $\AA\cap\C[x]$ contains sufficiently large
powers of the ideal $\mathfrak{m}_q$.

As in dimension $n=1$, we apply to $M$ the formal Fourier transform:
$M\to M^{\vee}$ and similarly denote by $\AA^{\vee}$,
$\mathfrak{m}_q^{\vee}$, $\R_e^{\vee},\ldots$ the objects dual to $\AA$,
$\mathfrak{m}_q$, $\R_e,\ldots$ under the corresponding Fourier
involution\footnote{Here, $\mathfrak{m}_q$, $\R_e$,$\ldots$ are
regarded as being imbedded in the Weyl algebra $A_n(\C)$.}.  Again, since $\Ord\AA=0$, it is clear that
$\AA^{\vee}\cap\C[\partial]$ is not empty.  Furthermore, $\AA^{\vee}$ is
a \textit{right} module over the (commutative) ring $\R_e^{\vee}$.  

Let $X_{\tau}$ be the open (quasi)-affine algebraic variety in $\C^n$
obtained by removing the zero set of the polynomial $\tau(x)$, i.e.
$X_{\tau}:=\C^n\backslash \tau^{-1}(0)$.  Then 
(see Example~\ref{examp:2}, Sect.~\ref{sec:regdifops}), we can identify
the ring $\D(X_{\tau})$ of regular differential operators on
$X_{\tau}$ with the localization of the Weyl algebra at
$\tau=\tau(x)$, i.e.,
$\D_{\tau}:=\D(X_{\tau})\cong\C[x,\tau^{-1}]\otimes_{\C[x]} A_n(\C)$.
Since $\D\subset\D_{\tau}$, one can define the \textit{extension ideal}
$\AA_{\tau}^{\vee}$ of $\AA^{\vee}$ in this localization,
namely $\AA_{\tau}^{\vee}:=\D_{\tau}\langle\AA^{\vee}\rangle$.  The
following lemma clarifies the algebraic structure of
$\AA_{\tau}^{\vee}$.

\begin{lemma}\label{lem:ideal} $\AA_{\tau}^{\vee}$ is a principal left ideal in
$\D_{\tau}$.
\end{lemma}
\begin{proof}

By construction, $\AA^{\vee}$ is a left ideal in $\D$ annihilating the
principal generator $e^{\vee}:=\tau(x)+\D\mathfrak{m}_q^{\vee}$ of the
$\D$-module $M^{\vee}$.  Fix $N\in\Z_+$ such that $N>\deg q$, and
define $G(x,\partial):=\tau^N(x)\circ q(\partial)\circ
\tau(x)^{-1}\in\D_{\tau}$.  Then, it follows that
$G(x,\partial)\in\AA^{\vee}$.  Indeed, in view of the Weyl algebra
commutation relations one has $\ad_{\tau}^N[q(\partial)]=0$, where
$\ad_{\tau}^N$ stands for the $N$-iterated commutator with
$\tau=\tau(x)$.  Since 
$$
\ad_{\tau(x)}^N[q(\partial)]=\sum_{k=0}^N(-1)^k {N\choose k}
\tau(x)^{N-k}\circ q(\partial)\circ \tau(x)^k,$$
the expression $\tau(x)^N\circ q(\partial)$ is divisible by $\tau(x)$
from the right within $A_n(\C)$.  On the other hand, it is obvious
that $G(x,\partial)\circ \tau(x)\equiv 0 \ (\textup{mod}\  \D\mathfrak{m}_q^{\vee})$.

Put $D=D(x,\partial):=\tau(x)^{-N+1}\circ G(x,\partial)$, so that
$D\in\AA_{\tau}^{\vee}$ in view of the definition of
$\AA_{\tau}^{\vee}$.  Then, we claim that the ideal
$\AA_{\tau}^{\vee}$ is generated by $D$ in $\D_{\tau}$.  Indeed, if
$Q(x,\partial)\in\AA_{\tau}^{\vee}$, then $Q$ is factorizable as
$Q(x,\partial)=S(x,\partial)\circ T(x,\partial)$ with $T\in\AA^{\vee}$
and $S\in\D_{\tau}$.  By definition of $\AA^{\vee}$, we have
$T(x,\partial)\circ\tau(x)\equiv 0 \ (\textup{mod}\
\D\mathfrak{m}_q^{\vee})$, and hence $T(x,\partial)\circ\tau(x)=\tilde
T(x,\partial)\circ q(\partial)$ for some $\tilde T(x,\partial)\in\D$.
Multiplying both sides of the last equation by $S(x,\partial)$ on the
left and by $\tau(x)^{-1}$ on the right, we get $Q(x,\partial)=\tilde
T'(x,\partial)\circ D(x,\partial)$ where $\tilde T'=\tilde T\circ
\tau(x)^{-1}\in\D_{\tau}$.  Thus, $\AA_{\tau}^{\vee}$ is a cyclic left
ideal in $\D_{\tau}$.
\end{proof}

Now we are ready to state the main theorem which completes our
construction of algebraic Darboux transformations related to the
$\D$-module $\Delta(q)$.

\begin{theorem}\label{thm:multfree}
Associated with $M\subset\Delta(q)$ there exists a commutative ring
$\R_M\subset\D_{\tau}$ of regular differential operators on
$X_{\tau}=\C^n\backslash \tau^{-1}(0)$ isomorphic to $\R_e$.  If
$\Lambda(q)$ and $\Lambda(\tau)$ are the complex linealities\footnote{By a
\textit{complex lineality } of a polynomial $P\in\C[x]$ we mean the
maximal linear subspace $\Lambda(P)$ of $\C^n$ such that $P$ can be
restricted to a polynomial on the quotient $\C^n/\Lambda(P)$.} of the
polynomial $q$ and $\tau$ and if
$\Lambda(q)\cup\Lambda(\tau)\not=\C^n$, the ring $\R_M$ is nontrivial
in the sense that $\R_M$ cannot be transformed into
$\R_e^{\vee}\subset\C[\partial]$ by conjugation by a regular function
on $X_{\tau}$ or through a local nonsingular change of variables.
\end{theorem}

\begin{proof}
By definition, $\AA_{\tau}^{\vee}$ is generated by $\AA^{\vee}$ as a
\textit{left} module over $\D_{\tau}$.  Hence, it inherits (from
$\AA^{\vee}$) the structure of a \textit{right} module over the
commutative ring $\R_e^{\vee}$.  Therefore, in view of
Lemma~\ref{lem:ideal}, we conclude that for \textit{every}
$L_0\in\R_e^{\vee}$ there exists an operator $L\in\D_{\tau}$ such that
$D\circ L_0=L\circ D$ in $\AA_{\tau}^{\vee}$ with $D=\tau(x)\circ
q(\partial)\circ \tau(x)^{-1}$.  Denote by $\R_M$ the set of all
differential operators $L\in\D_{\tau}$ constructed through this
procedure.  Then, $\R_M$ forms a
\textit{commutative} subalgebra in the ring $\D(X_{\tau})$ since the
latter has no zero divisors.  When
$\Lambda(q)\cup\Lambda(\tau)\not=\C^n$,
$\Lambda^{\perp}(q)\cap\Lambda^{\perp}(\tau)$ is not empty and
$[q(\partial),\tau(x)]\not=0$, so that $D\not\in\C[\partial]$.  This
implies that $\R_M$ is neither equal nor conjugate to $\R_e^{\vee}$.
Moreover, since the operators in $\R_M$ have constant principal
symbols, they cannot be reduced to their principal parts through a
local change of coordinates. 
\end{proof}

\begin{remark}\label{rem}
Recall that, by construction, $\R_e$ (and hence $\R_e^{\vee}$) is not
empty.  In particular, $\R_e$ contains the powers
$\mathfrak{m}_q^N\supset \mathfrak{m}_q^{N+1}\supset\ldots$ of the
ideal $\mathfrak{m}_q$ starting with $N=\deg\tau(x)+1$.  By Fourier
duality, this implies that
$(\mathfrak{m}_q^{\vee})^N\subseteq\R_e^{\vee}$ for all
$N\geq\deg\tau(x)+1$.  In fact, if $L_0\in(\mathfrak{m}_q^{\vee})^N$,
then the corresponding operator $L\in \R_M$ can be constructed
explicitly.  Indeed, we have $L_0=p(\partial)\circ q(\partial)^N$ with
$p(\partial)\in\C[\partial]$.  On the other hand, in view of the Weyl
algebra commutation relations, the identity
$$
\ad_{q(\partial)}^N[\tau(x)]:=\left[q(\partial),\left[q(\partial),\ldots[q(\partial),\tau(x)]\ldots\right]\right]=0,
$$ or more explicitly, $$
\sum_{k=0}^N(-1)^k{N\choose k} q(\partial)^{N-k}\circ \tau(x)\circ
q(\partial)^k=0
$$
holds for all $N\geq \deg\tau+1$.  Whence, we get
$$
q(\partial)^N\circ\tau(x)=\sum_{k=1}^N(-1)^{k+1}{N \choose k}
q(\partial)^{N-k}\circ \tau\circ q(\partial)^k,
$$
so that $L_0$ can be factorized in the form 
\begin{equation}
L_0=Q(x,\partial)\circ D(x,\partial)\label{eqn:factors}.
\end{equation}
Here \begin{equation}Q(x,\partial):=\sum_{k=1}^N(-1)^{k+1}{N\choose k}
p(\partial)q(\partial)^{N-k}\circ \tau(x)\circ
q(\partial)^{k-1}\circ\tau(x)^{-1}\label{eqn:explicit}
\end{equation}
and $D(x,\partial):=\tau(x)\circ q(\partial)\circ\tau(x)^{-1}$.  By
interchanging the factors in \eqref{eqn:factors}, we obtain the
corresponding operator in $\R_M$: 
\begin{equation}
L:=D(x,\partial)\circ Q(x,\partial).\label{eqn:dt}
\end{equation}
  This procedure is a
\textit{precise} analog of the classical one-dimensional Darboux
transformation \cite{Darboux}.
\end{remark}

\subsubsection{The general case}  This procedure of algebraic Darboux
transformations can be extended to the case of arbitrary
multiplicities.  Again, we start with a pair of complex polynomials
$q=q(x)$ and $\tau=\tau(x)$ in $n\geq1$ independent variables (without
 assuming $q$ to be primitive).  In this case, we can decompose
$q=Cq_1^{k_1}q_2^{k_2}\ldots q_s^{k_s}$ into a product of powers of
pairwise distinct irreducible polynomials $q_i=q_i(x)\in\C[x]$ with
(strictly) positive integer multiplicities $k_i\in\Z_+$,
$C\in\C^{\times}$, and we write $\mathfrak{m}_i:=\mathfrak{m}_{q_i}$
for the corresponding prime ideals in $\C[x]$.  To each $k_i$ we
then associate the $\C$-linear vector space $\V_i\subset \C[x]$
spanned by all multiples of $\tau(x)$ by polynomials of order less
than $k_i$, i.e. we set 
$$
\V_i=\Span_{\C}\langle \alpha(x)\tau(x)\in\C[x] \ : \
\deg\alpha<k_i\rangle.
$$
With the notation of Section~\ref{sec:multfree} we define
$\Delta(q_1,\ldots,q_s):=\bigoplus_{i=1}^s \Delta(q_i)$ as a left
$\D$-module over the Weyl algebra on $\C^n$ and consider its submodule
$M$ generated by the elements $e_i:=p_i(-\partial)+\D\mathfrak{m_i}$
with all $p_i\in\V_i$, $i=1,\ldots,s$.

Let $M^{\vee}$ be a Fourier transform of the $\D$-module $M$.
Clearly, $M^{\vee}=\D\langle e_1^{\vee},\ldots,e_s^{\vee}\rangle$,
where $e_i^{\vee}:=\V_i+\D\mathfrak{m}_i^{\vee}$, while
$\mathfrak{m}_i^{\vee}$ stands for the image of $\mathfrak{m}_i$ under
the (formal) Fourier involution in $\D$.  We write
$\AA^{\vee}\subset\D$ for the left ideal in $\D$ annihilating the
elements $e_i^{\vee}$, $i=1,\ldots,s$,
$\AA^{\vee}:=\Ann_{\D}(e_1^{\vee},\ldots,e_s^{\vee})$ and consider its
extension $\AA_{\tau}^{\vee}:=\D_{\tau}\langle\AA^{\vee}\rangle$ in
the localization of the Weyl algebra $\D_{\tau}$ at $\tau$.  

In this setting, we can refine the argument leading to the proof of
Lemma~\ref{lem:ideal}, i.e. we can show the \textit{cyclicity} of the
ideal $\AA_{\tau}^{\vee}$ within $\D_{\tau}$.  Indeed, if
$Q(x,\partial)\in\AA_{\tau}^{\vee}$, one has
$Q(x,\partial)=S(x,\partial)\circ T(x,\partial)$ with
$S(x,\partial)\in\D_{\tau}$ and $T(x,\partial)\in\AA^{\vee}\subset\D$.
The latter implies that $T(x,\partial)\circ p(x)=\tilde
T_p(x,\partial)\circ q_i(\partial)$ for \textit{any} $p(x)\in\V_i$ and
some $\tilde T_p\in\D$.  Since $p(x)=\alpha(x)\tau(x)$ with
$\alpha(x)\in\C[x]$, $\deg \alpha<k_i$, we can rewrite the above
equation in the form $R(x,\partial)\circ\alpha(x)=\tilde T_p(x,\partial)
\circ q_i(\partial)$, where $R(x,\partial):=T(x,\partial)\circ
\tau(x)\in\D_{\tau}$.  It follows that all the derivatives of the
\textit{full} symbol $R(x,\xi)$ of the operator $R$ up to order
$k_i-1$ with respect to $\xi$ vanish on the zero set of the ideal
$\mathfrak{m}_i$ in $\C^n$.  Elementary induction in $k_i$ with the
use of the Nullstellensatz gives immediately that
$R(x,\xi)=R_i(x,\xi)q_i^{k_i}(\xi)$ and, hence, $T(x,\partial)$ is
divisible on the right by $q_i(\partial)^{k_i}\circ\tau(x)^{-1}$ for
all $i=1,2,\ldots,s$.  Finally, due to the commutativity of the
operators $D_i:=\tau(x)\circ q_i(\partial)^{k_i}\circ\tau(x)^{-1}\in
\D_{\tau}$, we conclude that $T$ must be of the
form $T(x,\partial)=\tilde T'(x,\partial)\circ D(x,\partial)$ where
$D:=\prod_iD_i=\tau(x)\circ q(\partial)\circ \tau(x)^{-1}$ and $\tilde
T'\in\D_{\tau}$.  In this way, the operator $D(x,\partial)$ is a
principal generator of the ideal $\AA_{\tau}^{\vee}$ in $\D_{\tau}$.  

Let $\R_e:=\{ P\in\C[x]\ :\ P\,.\,\Span_{\C}(e_1,\ldots,e_s)\subseteq
\Span_{\C}(e_1,\ldots e_s)\}$ and let $\R_e^{\vee}$ be its
formal Fourier transform.  Both $\R_e$ and $\R_e^{\vee}$ are not
empty: they contain in particular the ideals $\cap\,
\mathfrak{m}_i^{k_i}$ and $\cap(\mathfrak{m}_i^{\vee})^{k_i}$
respectively.  Then, the left ideals $\AA^{\vee}$ and
$\AA_{\tau}^{\vee}$ have the structure of right modules over the
commutative ring $\R_e^{\vee}$, and therefore we again arrive at the
result stated in Theorem~\ref{thm:multfree}.  Namely, we obtain a
commutative ring $\R_M\subset \D_{\tau}$ of regular differential
operators on $X_{\tau}=\C^n\backslash\tau^{-1}(0)$ isomorphic to
$R_e^{\vee}$.  This completes the procedure of algebraic Darboux
transformations in the case of arbitrary multiplicities.

\subsection{The Bispectral Duality}

The formal spectral problem for the commutative ring $\R_M$ has a
unique solution of a particularly simple structure.  Indeed, by
construction, the (quasi-)regular function on $T^*X_{\tau}$ given by
$\psi=\psi(x,\xi):=D(x,\partial)e^{(x,\xi)}$ is a common eigenfunction
for all differential operators in $\R_M$.  We call $\psi$ the
\textit{Baker-Akhiezer function} associated with $\R_M$.  

The following elementary argument shows that there exists a certain
complete commutative ring $\R_M^{\flat}$ of partial differential
operators on $X_q=\C^n\backslash q^{-1}(0)$ written in terms
of the spectral variable $\xi$ for which $\psi$ is also a (unique)
common eigenfunction.

Put $N=\deg q(x)$ and denote by $I^{\flat}$ the principal ideal in the
polynomial ring $\C[x]$ generated by $\tau(x)^N$.  For every
$\gamma\in I^{\flat}$ we have
$\gamma(x)\psi(x,\xi)=Q_{\gamma}(x,\partial_x)e^{(x,\xi)}$ with some
$Q_{\gamma}\in A_n(\C)$.  Then, there obviously exists another
differential operator
$Q_{\gamma}^{\flat}(\xi,\partial_{\xi})$ in $A_n(\C)$ written formally
in terms of $\xi$ such that
$Q_{\gamma}(x,\partial_x)e^{(x,\xi)}=Q_{\gamma}^{\flat}(\xi,\partial_{\xi})e^{(x,\xi)}$,
and hence
\begin{equation}
Q_{\gamma}^{\flat}(\xi,\partial_\xi)e^{(x,\xi)}=\gamma(x)\psi(x,\xi)\label{eqn:gamma}.
\end{equation}
On the other hand, we have 
\begin{equation}
\tau(\partial_{\xi})\psi(x,\xi)=\tau(x)q(\partial_x)e^{(x,\xi)}
=q(\xi)\tau(x)e^{(x,\xi)}\label{eqn:gamma2}
\end{equation}
in view of the definition of $\psi$.  It follows immediately from
\eqref{eqn:gamma} and \eqref{eqn:gamma2} that
\begin{equation}
L_{\gamma}(\xi,\partial_\xi)\psi(x,\xi)=\gamma(x)\tau(x)\psi(x,\xi),\label{eqn:gamma3}
\end{equation}
where
$L_{\gamma}(\xi,\partial_{\xi}):=Q_{\gamma}^{\flat}(\xi,\partial_{\xi})\circ
q(\xi)^{-1}\circ \tau(\partial_{\xi})$.  The family of all operators
$L_{\gamma}(\xi,\partial_{\xi})$ constructed in this way forms a
commutative ring $\R_M^{\flat}$ of partial differential operators in
the localized Weyl algebra $\D_{q}\cong
\C[\xi,q(\xi)^{-1}]\otimes_{\C[\xi]}A_n(\C)$ isomorphic to $I^{\flat}$.  This ring can be regarded as a
\textit{bispectral dual} of $\R_M$ in the sense of \cite{DGr,W1}.  The
simple procedure we have applied to generate the ring $\R_M^{\flat}$
is, in fact, an extension of the bispectral Darboux transformation
scheme in $n=1$ as discussed in \cite{KR,Kbdt}.

In conclusion, we note that the ring $\R_M^{\flat}$, though complete,
is not necessarily a \textit{maximal commutative} subring in $\D_{q}$,
i.e., the centralizer $Z(\R_M^{\flat})$ of $\R_M^{\flat}$ may happen
to be larger than $\R_M^{\flat}$ itself.  It is an interesting problem
to clarify the structure of $Z(\R_M^{\flat})$ in general.

\section{Examples, Applications and Concluding Remarks}\label{sec:examples}

The purpose of this section is to display some explicit examples of
the theory presented so far and to illuminate some of its
implications.

Let $n=2$.  We choose $\tau(x_1,x_2)=x_1^2-x_2$ and
$q(x_1,x_2)=x_1x_2-\lambda$, $\lambda\in\C$.  If $\lambda\not=0$, the
polynomial $q$ is irreducible.  According to
Theorem~\ref{thm:multfree}, for any constant coefficient operator
$Q\in\C[\partial_1,\partial_2]$ and
$L_0=(\partial_1\partial_2-\lambda)^3$ there exists a unique partial
differential operator $L_Q$ satisfying the identity
\begin{equation}
D\circ Q\circ L_0=L_Q\circ D,\label{eqn:exampshift}
\end{equation}
where
\begin{multline}
D:=\tau(x_1,x_2)\circ
q(\partial_1,\partial_2)\circ\tau(x_1,x_2)^{-1}\\
=
\left(\partial_1-\frac{2x_1}{x_1^2-x_2}\right)\left(\partial_2+\frac{1}{x_1^2-x_2}\right)-\lambda.
\end{multline}
For example, if $Q\equiv1$, formulas \eqref{eqn:factors}--\eqref{eqn:dt} (see Remark~\ref{rem}) give explicitly
\begin{multline} L_1 = {\partial_1^3}{\partial_2^3} +
{\frac{-6\left( {x_1^2} + x_2 \right) }{{{\left( {x_1^2} - x_2 \right)
}^2}}} \relax \partial_1{\partial_2^3} + {\frac{24x_1x_2}{{{\left(
{x_1^2} - x_2 \right) }^3}}} \relax {\partial_2^3} \\
+
{\frac{-3\left( -4x_1 + \lambda{x_1^4} - 2\lambda{x_1^2}x_2 +
\lambda{x_2^2} \right) } {{{\left( {x_1^2} - x_2 \right) }^2}}}
\relax {\partial_1^2}{\partial_2^2}
 + {\frac{-6\left(
13{x_1^2} + 5x_2 \right) } {{{\left( {x_1^2} - x_2 \right) }^3}}}
\relax \partial_1{\partial_2^2}\\
 + {\frac{6\left( 18{x_1^3} +
\lambda{x_1^6} + 30x_1x_2 - \lambda{x_1^4}x_2 - \lambda{x_1^2}{x_2^2}
+ \lambda{x_2^3} \right) }{{{\left( {x_1^2} - x_2 \right) }^4}}}
\relax {\partial_2^2} + {\frac{-3}{{{\left( {x_1^2} - x_2 \right)
}^2}}} \relax {\partial_1^3}\partial_2 \\
+ {\frac{54x_1}{{{\left(
{x_1^2} - x_2 \right) }^3}}} \relax {\partial_1^2}\partial_2+
{\frac{3\left(6{{\lambda}^2}{x_1^4}{x_2^2} - 
4{{\lambda}^2}{x_1^2}{x_2^3} + {{\lambda}^2}{x_2^4} \right) }{{{\left(
{x_1^2} - x_2 \right) }^ 4}}} \relax \partial_1\partial_2 \\
+ {\frac{3\left( -118{x_1^2} - 4\lambda{x_1^5} + {{\lambda}^2}{x_1^8}
- 26x_2 + 8\lambda{x_1^3}x_2 - 4{{\lambda}^2}{x_1^6}x_2 - 4\lambda
x_1{x_2^2} \right) }{{{\left(
{x_1^2} - x_2 \right) }^ 4}}} \relax \partial_1\partial_2 
\\
+{\frac{6\left( 136{x_1^3} + 9\lambda{x_1^6} + 104x_1x_2 -
15\lambda{x_1^4}x_2 + 3\lambda{x_1^2}{x_2^2} + 3\lambda{x_2^3} \right)
}{{{\left( {x_1^2} - x_2 \right) }^5}} } \relax \partial_2 +
{\frac{-3}{{{\left( {x_1^2} - x_2 \right) }^3}}} \relax
{\partial_1^3}\\
 + {\frac{3\left( 24x_1 + \lambda{x_1^4} -
2\lambda{x_1^2}x_2 + \lambda{x_2^2} \right) } {{{\left( {x_1^2} - x_2
\right) }^4}}} \relax {\partial_1^2}\\
 + {\frac{-12\left(
52{x_1^2} + 3\lambda{x_1^5} + 8x_2 - 6\lambda{x_1^3}x_2 + 3\lambda
x_1{x_2^2} \right) }{{{\left( {x_1^2} - x_2 \right) }^5}}} \relax
\partial_1\\
 + \frac{1920{x_1^3} - {{\lambda}^3}{x_1^{12}} +
960x_1x_2 + 6{{\lambda}^3}{x_1^{10}}x_2 -
15{{\lambda}^3}{x_1^8}{x_2^2} + 24\lambda{x_2^3}}{{{{\left( {x_1^ 2} -
x_2 \right) }^6}}} \\
+\frac{ - {{\lambda}^3}{x_2^6} +
20\lambda{x_1^6} \left( 6 + {{\lambda}^2}{x_2^3} \right) +
6\lambda{x_1^2}{x_2^2} \left( 12 + {{\lambda}^2}{x_2^3} \right) -
3\lambda{x_1^4}x_2 \left( 72 + 5{{\lambda}^2}{x_2^3} \right)
}{{{\left( {x_1^ 2} - x_2 \right) }^6}}. \label{eqn:examp1}
\end{multline} 
Note that the operators $L_Q$ are nontrivial in the sense that they
cannot be reduced to the constant operators $L_Q^0:=Q\circ L_0$ by means of
conjugation or change of variables.

The simplest operator in the spectral parameter which has the same
eigenfunction as the $L_Q$'s can also be computed in an explicit form
using \eqref{eqn:gamma}, \eqref{eqn:gamma2} and \eqref{eqn:gamma3}.
Namely, we have 
\begin{equation}
L_1^{\flat}(\xi,\partial_{\xi})=Q_1^{\flat}\circ
q(\xi_1,\xi_2)^{-1}\circ \tau(\partial_{\xi}),\label{eqn:examp2}
\end{equation}
where 
\begin{multline*}
Q_1^{\flat}(\xi,\partial_{\xi})=
(\xi_1\xi_2-\lambda)\partial_{\xi_1}^4
-2\xi_2 \partial_{\xi_1}^3
-2(\xi_1\xi_2-\lambda) \partial_{\xi_1}^2\partial_{\xi_2}\\
+\xi_1 \partial_{\xi_1}^2
+(\xi_1\xi_2-\lambda) \partial_{\xi_2}^2
+2\xi_2 \partial_{\xi_1}\partial_{\xi_2}
-\xi_1 \partial_{\xi_2}
-4\partial_{\xi_1}.
\end{multline*}

The next example reveals the link to the theory of Huygens' principle
on Minkowski spaces.  Let $n$ be an arbitrary even number, $n\geq 4$.
Take  $\tau(x)=x_1^2-x_2$ and
$q(x)=x_1x_2-x_3^2-x_4^2-\cdots-x_n^2$.  Clearly, $q$ is again an
irreducible polynomial and we can apply Theorem~\ref{thm:multfree}.
The simplest operator in the commutative ring $\R_M$ has the same form
as \eqref{eqn:examp1}, where the parameter $\lambda$ is  formally
replaced by the Laplacian
$\Delta_{n-2}=\partial_3^2+\partial_4^2+\cdots+\partial_n^2$ in $n-2$
extra variables.  This makes sense since $\lambda$ enters polynomially
in to $L_0$ and $L_1$.  The operator $\tilde
L_0=(\partial_1\partial_2-\partial_3^2-\cdots\partial_n^3)^3$ is a
cube of the classical wave operator, and hence, it satisfies
Huygens' principle for all \textit{even} $n\geq 8$ (cf., e.g., \cite{Riesz}).

Adapting the technique developed in \cite{Ber3}, we can prove that 
the operator $\tilde L_1=\partial_1^3\partial_2^3+\ldots$ obtained by
the formal substitution $\lambda\mapsto\Delta_{n-2}$ in \eqref{eqn:examp1} is also a
nontrivial Huygensian hyperbolic operator in even dimensions $n\geq 12$.
Note that  $\tilde L_1$ \textit{cannot} be presented as a
power of the second order wave-type operator on the Minkowski space
$\mathbb{M}^n$.  Therefore, we get a new interesting example related to
the classical Hadamard's problem (see \cite{BV2,Had} and
references therein).




\end{document}